\documentclass[aps,prd,reprint,nofootinbib,amsmath,amssymb]{revtex4-2}

\usepackage{graphicx}
\usepackage{dcolumn}
\usepackage{bm}
\usepackage{hyperref}
\usepackage{color}
\usepackage{physics}

\newcommand{\dif}{\mathrm{d}}

\begin{document}

\title{Kerr Perihelion Precession via the Laplace-Runge-Lenz Vector Method}

\author{Sidney Natzuka Junior}
\email{sidney.natzuka@unesp.br}
\affiliation{Instituto de Física Teórica, Universidade Estadual Paulista (UNESP), \\
Rua Dr. Bento Teobaldo Ferraz, 271, Bloco II -- Barra Funda, 01140-070, São Paulo, SP, Brazil}

\date{December 29, 2025}

\begin{abstract}
We calculate, up to the first-order in the black hole spin, the perihelion precession of a test particle in the equatorial plane of a Kerr black hole using the perturbative Laplace-Runge-Lenz (LRL) vector method. To account for the dragging of inertial frames, we modify the LRL vector by incorporating a counteracting term in the angular momentum, which preserves the Keplerian orbit form to first order. We derive the standard Lense-Thirring precession result, leading to a transparent reinterpretation of known results, clarifying the role of frame dragging in LRL-based perturbation methods.
\end{abstract}

\maketitle

\section{Introduction}

The perihelion precession of planetary orbits serves as a fundamental test of general relativity. While the Schwarzschild contribution $\Delta \omega_{S} = 6\pi M/p$ is standard textbook material \cite{hartle,mtw}, the influence of the central body's rotation (Kerr metric) introduces corrections for the angular velocity, stated as the Lense-Thirring effect \cite{lammerzahl,iorio2}. The standard result for the perihelion precession due to the central body's angular momentum $J=Ma$ in the equatorial plane is given by $\Delta \omega_{LT} \approx  -8\pi J / (p^{3/2} M^{1/2})$ \cite{landau,weinberg,brill1999}. Additionally, spin couplings and rotation effects can serve as experimental verification for fundamental theories, such as in general relativity with the Lense-Thirring effect. Past results like the Gravity Probe B mission \cite{everitt} have shown close agreement with the theory, while new measurements can be further tests for the theory \cite{pasham}.

Standard derivations of the perihelion precession often rely on the effective potential method, which necessitates the evaluation of elliptic integrals or Hamilton-Jacobi equations \cite{kraniotis,chakraborty, hu, dean, bambhaniya, heydarifard, he24, andreoli}. While rigorous, these methods can obscure the physical origin of specific perturbative terms. In contrast, this work employs the LRL perturbation method, which has been recently used in different scenarios to explore the dynamics in a more illustrative manner \cite{batic, silenko, xu2011}. This approach offers a different interpretation of the dynamics, allowing us to isolate and identify the specific physical mechanisms—such as the coupling between frame-dragging and radial acceleration—that drive the precession. Pedagogically, this aligns with efforts to illustrate dynamical symmetries using vector methods rather than purely integration \cite{oconnell}.

A critical subtlety often handled via modified LRL vector definitions or magnetic analogies \cite{brill1999, chashchina} is the definition of the angular velocity $\dot{\phi}$ within the perturbative integral. In the Kerr metric, $\dot{\phi}$ deviates from the Keplerian value $L/r^2$ due to the frame-dragging effect of Kerr spacetime. We show that to obtain correct first-order results in the spin parameter $a$, a modification in the LRL vector must be introduced to avoid first-order corrections in the orbit.

\section{The Unperturbed Kepler Problem}\label{sec:ukp}

We begin by establishing the reference Newtonian system. Due to angular momentum conservation, the motion around a central mass is restricted to a plane perpendicular to the angular momentum. Hence, we define the axis in the following way: the angular momentum points to the $\vu e_z$ direction, so the motion is contained in the $xy-$plane. The $\vu e_x$ direction points from the central mass to the perihelion, and the $\vu e_y$ direction follows from the right-hand rule. Furthermore, there is no need to define the time in this reference frame, due to the universality of the Newtonian time; however, when establishing a connection with the Kerr metric in Sec \ref{sec:lrl_pert}, this coordinate time will be the proper time of the test mass. Throughout the text, we will make use of polar ($r,\phi$) coordinates in the $xy-$plane, defined by $x=r\cos \phi$ and $y=r\sin \phi$. 
Consider a particle of mass $m=1$ (we set the test particle mass to unity for simplicity) moving in a Euclidean space under a central potential $V(r) = -M/r$, where $M$ is the mass of the central body. The equation of motion is:
\begin{equation}
    \ddot{\vb{r}} = -\frac{M}{r^2}\vu{e}_r.
\end{equation}
The unperturbed orbit is described by the conservation of energy (per unit mass) $E_0$ and specific angular momentum (per unit mass) $\vb{L} = \vb{r} \times \vb{v} = L\, \vu{e}_z$. The LRL vector, defined as:
\begin{equation}\label{eq:lrl}
    \vb{A} = \vb{v} \times \vb{L}_K - M\vu{e}_r,
\end{equation}
is a constant of motion for this unperturbed system ($\dif \vb{A}/\dif t = 0$ strictly for the $1/r$ potential), reflecting the hidden $SO(4)$ symmetry of the Kepler problem \cite{goldstein, bander1966, leach2003}. Its magnitude is $A = eM$. We can recover the orbit by taking the dot product of the LRL vector with the radial vector:
\begin{align}\label{eq:lrl_eq1}
    \vb A \cdot \vb r &= \qty(\vb v \times \vb L)\cdot \vb r -  M \vb r \cdot \vu e_r
\end{align}

But $\vb A \cdot \vb r = Ar \cos \phi$, and $\qty(\vb v \times \vb L)\cdot \vb r= \qty(\vb r \times \vb v)\cdot \vb L=L^2$, then
\begin{equation}\label{eq:conic_sol}
    Ar \cos \phi = L^2 - Mr\epsilon \Rightarrow r(\phi) = \frac{L^2/M}{1+(A/M)\cos \phi}.
\end{equation} 
The LRL vector presents a way of finding the solution to the Kepler problem without the necessity of solving a differential equation. The above solution is a conic with eccentricity $A/M$ (so $A=eM$) and semi-latus rectum $p=L^2/M$. By choosing a specific point on the orbit, it can be shown that $\vb A = A \vu{e}_x$; hence, the LRL vector points from the central mass to the perihelion.

\section{Geodesic Equations in Kerr Metric}

We now consider the full relativistic system: the Kerr metric in Boyer-Lindquist coordinates. We adopt geometric units where $G=c=1$. For the original derivation, see \cite{kerr1963}, for a review of its properties and consequences see \cite{mtw,bardeen72,visser2007, adamo2014}. Note that in these units, mass $M$ and spin parameter $a$ (where $J=Ma$ is the black hole's angular momentum) have dimensions of length $[L]$, while the specific angular momentum of the test particle $L$ also has dimensions of length $[L]$ (since $L \sim rv/c$). The specific energy $E$ is dimensionless. 

For a particle restricted to the equatorial plane ($\theta = \pi/2$), the metric is \cite{bardeen72}:
\begin{align}
    \dif s^2 = &-\left(1-\frac{2M}{r}\right)\dif t^2 - \frac{4Ma}{r} \dif t \dif \phi + \frac{r^2}{\Delta}\dif r^2 \nonumber \\
    & + \left(r^2+a^2+\frac{2Ma^2}{r}\right) \dif \phi^2,\label{eq:kerr_metric}
\end{align}
where $\Delta = r^2 - 2Mr + a^2$. The symmetries of the spacetime imply two constants of motion: the specific energy $E = -p_t$ and the angular momentum $L = p_\phi$.

It is worth noticing that these relativistic constants are not the same as their Newtonian counterparts. While $L$ and $E$ are exact constants of the motion in the Kerr spacetime, the quantities $L_K$ and $E_K$ used in the unperturbed orbit approximation are related but distinct. In the weak-field limit, $L \approx L_K$ and $E \approx 1 + E_K$ (where the rest mass energy is included in $E$). 

Inverting the momentum relations $p_\mu = g_{\mu\nu}\dot{x}^\nu$ yields the coordinate velocities with respect to proper time $\tau$. The angular velocity $\dot{\phi} \equiv d\phi/d\tau$ is given by:
\begin{equation} \label{eq:phi_dot_exact}
    \dot{\phi} = \frac{1}{\Delta}\left[\left(1-\frac{2M}{r}\right)L + \frac{2Ma}{ r}E\right].
\end{equation}
This expression differs from the Newtonian form by terms of order $\mathcal{O}(Ma/r)$ and higher. The term proportional to $aE$ represents the dragging of inertial frames \cite{bardeen72}. A similar equation for $\dot t$ is also obtained:
\begin{equation}
    \dot t = \frac{1}{\Delta}\left(\left(r^2+a^2\left(1+\frac{2M}{r}\right)\right)E-\frac{2Ma}{r}L\right)
\end{equation}
Using the normalization condition $g_{\mu\nu}\dot{x}^\mu \dot{x}^\nu = -1$, the radial equation of motion is derived as:
\begin{equation}\label{eq:dotr}
    \dot{r}^2 = E^2 - 1 + \frac{2M}{r} - \frac{L^2-a^2(E^2-1)}{r^2} + \frac{2M(L-aE)^2}{r^3}.
\end{equation}
Differentiating with respect to proper time $\tau$ leads to the radial acceleration:
\begin{equation}\label{eq:ddotr}
    \ddot{r} = -\frac{M}{r^2} + \frac{L^2-a^2(E^2-1)}{r^3} - \frac{3M}{r^4}(L-aE)^2.
\end{equation}

\section{Frame-Dragging and the Modified LRL Vector}\label{sec:lrl_pert}

Since the LRL vector is directed towards the perihelion, as stated at the end of Sec. \ref{sec:ukp}, a measure of perihelion shift can be obtained by computing the perpendicular variation of the LRL along an orbit. The simplicity of the method lies in calculating this variation on the original Newtonian orbit, considering that relativistic corrections do not substantially modify the test mass motion, hence this is a suitable approximation for far orbits ($M/r\ll 1$ and $a/r\ll 1$). Throughout this section we retain terms up to first order in $M/r$ and $a/r$, neglecting $\mathcal{O}(a^2/r^2)$, $\mathcal{O}(M^2/r^2)$, and mixed higher-order terms.

To utilize the LRL vector perturbation method, we reinterpret the geodesics in the Kerr spacetime as a motion under an effective force in a flat Euclidean spacetime. The time coordinate in this auxiliary space is identified with the proper time $\tau$ of the particle, such that the Newtonian equations of motion $\ddot{\vb{r}} = \dots$ refer to derivatives with respect to $\tau$. Even though we could define another time coordinate, the calculated perihelion precession will be the same, hence we choose the time coordinate in the easiest way possible. The mapping of the curved spacetime trajectory $r(\phi)$ onto the flat space Kepler orbit $r_0(\phi)$ introduces errors of higher order in the expansion parameter $M/r$, which are consistent with our perturbative approach.

However, since the LRL vector is related to the angular momentum (see Eq. \eqref{eq:lrl}), we must have a consistent description of this quantity when applying the mapping. If the spacetime were static, such as in the Schwarzschild metric, then the angular momentum would be functionally the same ($L \sim r^2 \dot \phi$). However, the Kerr metric (Eq. \eqref{eq:kerr_metric}) has a non-zero metric component $g_{t\phi}$ associated with the spacetime rotation, and therefore the angular momentum incorporates this rotation effect. From Eq. \eqref{eq:phi_dot_exact} we may write
\begin{align}
    L &= r^2 \dot \phi - \frac{2EMa}{r} + \mathcal{O}(M^2/r^2,a^2/r^2)\\
    \dot \phi &= \frac{L}{r^2} + \frac{2EMa}{r^3}+ \mathcal{O}(M^2/r^2,a^2/r^2).\label{eq:dot_phi_approx}
\end{align}
In order to achieve such consistency, we modify the angular momentum in the LRL vector in order to satisfy the following two conditions: (1) - it must be the same as the original LRL vector in Eq. \eqref{eq:lrl} in the limit of $a \to 0$, that is, when the canonical angular momentum matches the keplerian one, (2) - the orbit equation should only have a second order correction in the factor $a$, so first order corrections on the perihelion precession do not need to take into account the orbit change. Physically, changing the value of the angular momentum in the LRL vector is the same as changing the angular coordinate by $\phi'= \phi - \omega t$, and if $\omega$ is set to such a value that counteract the frame dragging effect, then at first order the orbit solution is the conic found in Eq. \eqref{eq:conic_sol}. The frame-dragging effect can be isolated from the previous equation:
\begin{align}
    L = r^2 \dot \phi' \Rightarrow \phi' = \phi - \frac{2Ma}{r^3}t.
\end{align}
The angular velocity $\omega = 2Ma/r^3$ is a first-order representation of the frame dragging. For a ZAMO (Zero Angular Momentum Observer), $\omega$ is the angular velocity of its trajectory. This gives a clear picture of what frame-dragging is: even ZAMOs have an angular velocity due to spacetime rotation. Then, a suitable redefinition of the LRL vector is
\begin{equation}\label{eq:modified_lrl}
    \vb A = \vb v \times \vu e_z \qty(L -\frac{2  EM a}{r})-  M \vu e_r.
\end{equation}
This specific form of the modified LRL vector was previously utilized by Brill and Goel to obtain light bending and precession for the Lense-Thirring metric \cite{brill1999}. Here, we provide a physical justification for this modification: the term $-2EMa/r$ acts as a counter-rotation to the Lense-Thirring (or Kerr in first order) frame dragging, effectively mapping the dynamics back to a quasi-Newtonian frame where the standard LRL conservation holds (to first order). By a similar procedure as in Eqs. \eqref{eq:lrl_eq1} and \eqref{eq:conic_sol}, but using that $\vb r \times \vb v = r^2 \dot \phi\, \vu e_z$, the orbit solution from this LRL vector is
\begin{equation}
    r(\phi) = \frac{1}{1+(A/M)\cos \phi}\qty(\frac{L^2}{M}-\frac{4E^2M a^2}{r^3}).
\end{equation}
As we can see from the above equation, further corrections to the Keplerian orbit are in the order $a^2$ by this modified LRL vector. For a first-order result, the contribution of $a$ for the orbit equation can be ignored. This would not be the case if the LRL vector were not modified (or modified differently). However, it still makes sense to believe that the LRL method could be applied successfully in these other cases if orbit corrections were taken into account.

\section{Calculation of the Perihelion Precession}\label{sec:calculation}

In this section, we will obtain the perihelion precession from the modified LRL vector given by Eq. \eqref{eq:modified_lrl}. Let $\vb v = \dot r \vu e_r +r\dot \phi \vu e_{\phi}$:
\begin{align}
    \vb A &= \qty(\dot r \vu e_r + r \dot \phi \vu e_{\phi})\times \vu e_z \qty(L-\frac{2EMa}{r})- M \vu e_r \nonumber \\
    &= -\dot r\qty(L-\frac{2EMa}{r})\vu e_{\phi}+\qty(r\dot \phi \qty(L-\frac{2EMa}{r})- M)\vu e_r
\end{align}
By substituting $\dot \phi$ and neglecting second order terms ($a^2$) we have
\begin{equation}\label{eq:mod_lrl}
    \vb A = \qty(\frac{L^2}{r}- M)\vu e_r - \dot r\qty(L-\frac{2EMa}{r})\vu e_{\phi}.
\end{equation}
The full derivation of the time variation of this modified LRL vector is done in Appendix \ref{sec:app_a}. Calculations have shown that in the leading order, only the variation of $\vb A$ in the $\phi$ direction is relevant. The result is presented:
\begin{equation}\label{eq:dot_a_phi}
    \dot{A}_{\phi}= \frac{3ML^3}{r^4}-\frac{8 EM^2a}{r^3}-\frac{2MEa(E^2-1)}{r^2}+\frac{6M^2L^2Ea}{r^5}
\end{equation}
The change of perihelion angle for a single orbit $\alpha$  can be obtained through an integration of the perpendicular component of the LRL vector velocity:
\begin{align}
    \alpha &= \int_{0}^{T} \dif t\, \frac{\qty(\vb A \times  \dot{\vb A})\cdot \vu e_z}{A^2},\nonumber \\
    &= \int_{0}^{2\pi} \frac{\dif \phi}{\dot \phi} \frac{\dot A_{\phi}\cos \phi}{A}.
\end{align}
Dividing $\dot A_{\phi}$ in Eq. \eqref{eq:dot_a_phi} by $\dot \phi$ in Eq. \eqref{eq:dot_phi_approx} yields
\begin{equation}
    \frac{\dot{ A}_{\phi}}{\dot \phi}= \frac{3ML^2}{r^2}-\frac{8 M^2Ea}{Lr}-\frac{2EM a(E^2-1)}{L}
\end{equation}
If we define
\begin{equation}
    I_n = \int_{0}^{2\pi} \dif \phi \frac{\cos \phi}{r^n}
\end{equation}
Then
\begin{equation}\label{eq:precession_angle}
    \alpha = \frac{1}{A}\qty[3ML^2 I_2-\frac{8M^2 E a}{L}I_1]
\end{equation}
Computing the integrals with respect to the unperturbed orbit (Eq. \eqref{eq:conic_sol}):
\begin{align}
    I_1 &= \frac{M}{L^2} \int_0^{2\pi} \cos\phi \qty(1+\frac{A}{M}\cos\phi) \dif \phi = \frac{A\pi}{L^2},\\
    I_2 &= \frac{M^2}{L^4} \int_0^{2\pi} \cos\phi \qty(1+\frac{A}{M}\cos\phi)^2 \dif \phi = \frac{2MA\pi}{L^4}.
\end{align}
Substituting these results in Eq. \eqref{eq:precession_angle} yields
\begin{equation}
    \alpha = \frac{6\pi M^2}{L^2}-\frac{8\pi M^2 E a}{L^3}
\end{equation}
The above result matches the perihelion precession obtained directly by the Lense-Thirring metric, as an effect of the rotation of the central body to the spacetime itself. By substituting $p=L^2/M$, $J=Ma$, and $E \approx 1$ (for non-relativistic far orbits with $M/r\ll 1$), we obtain the usual way of expressing the perihelion advance:
\begin{equation}
    \alpha = \frac{6\pi M}{p}- \frac{8\pi J}{p^{3/2}M^{1/2}}
\end{equation}
If $J=0$ ($a=0$), we reobtain the Schwarzschild perihelion precession. It is noticeable that, while the Schwarzschild contribution is positive, the Lense-Thirring is negative. It is possible to explain intuitively the positivity for the Schwarzschild contribution by imagining the gravitational potential as a sink for which the test particle is avoiding to fall due to its rotational velocity. The Schwarzschild correction to the potential is a factor of $2ML^2/r^3$ (see Eq. \eqref{eq:dotr}), then the sink steepness increases (relative to the Newtonian potential) as the test particle gets closer to the center (in the perihelion). Comparing with the Newtonian orbit, the perihelion and its vicinity get a little closer to $r=0$, thus by conservation of angular momentum, its angular velocity increases, dragging the perihelion forward by the calculated amount.

The black hole spin effect, however, can not be explained by a potential analogy with a sink, due to its simultaneous effect on both the potential and the angular velocity. It is more appropriate to interpret it as a spin-orbit coupling, which has a repulsive effect if the spin and the angular momentum are aligned. In our case, aligned means that the directions of the test mass angular momentum $\vb L = L \vu e_z$ and the black hole's angular momentum $\vb J = J \vu e_z$ are the same. If $J>0$ ($a>0$), we have repulsion and therefore the combined effect on the orbit is a slowdown yielding the perihelion shift retrogradely. The reverse happens if $J<0$ ($a<0$).

To quantify the magnitude of these effects in a real scenario, we consider the solar system parameters discussed in Ref. \cite{iorio2}. For Mercury, the Schwarzschild correction yields $5.01 \times 10^{-7}$ radians for each orbit, which corresponds to $43$ arcseconds per century. The Lense-Thirring effect goes with $p^{-3/2}$; consequently, in the solar system, the effect is larger also for Mercury, and the calculations yield a value $-2.38 \times 10^{-11}$ radians per orbit, or $-0.002$ arcseconds per century. According to current MESSENGER mission measurements \cite{park2017}, the Lense-Thirring precession is detectable with 75\% accuracy level.
\section{Conclusion}

We have derived the perihelion precession for a Kerr black hole using the LRL vector method. By keeping the first-order contribution of the black hole spin both in the perturbative force and the angular velocity $\dot{\phi}$, we accounted for the frame-dragging effect on the orbital frequency. Our perturbative approach yields a result consistent with the standard Lense-Thirring perihelion precession result $\alpha_{LT} \propto 8\pi J / p^{3/2}$ with the correct coefficient. The agreement confirms that the coupling between the radial perturbation on the effective force (or potential) and the angular velocity is necessary for the correctness of the perturbative description. Furthermore, this method can be extended to illustrate the kinematics in other spacetime geometries.

\begin{acknowledgments} The author thanks Professors C. A. D. Zarro and C. Farina for the helpful and inspiring discussions. This study was financed in part by the Coordenação de Aperfeiçoamento de Pessoal de Nível Superior - Brasil (CAPES) - Finance Code 001, grant number 88887.142182/2025-00. \end{acknowledgments}

\appendix

\section{Explicit Derivation of $\dot{A}_\phi$}\label{sec:app_a}
Here we detail the calculation of $\dot{\mathbf{A}}$ stated in Section \ref{sec:lrl_pert}. Starting with the modified LRL vector in Eq. \eqref{eq:mod_lrl}:
\begin{equation}
    \vb A = \qty(\frac{L^2}{r}- M)\vu e_r - \dot r\qty(L-\frac{2EMa}{r})\vu e_{\phi}.
\end{equation}
Differentiation with respect to proper time $\tau$ gives:
\begin{align}
    \dot{\mathbf{A}} &= \frac{d}{d\tau} \left[ \left(\frac{L^2}{r} - M\right)\mathbf{e}_r \right] - \frac{d}{d\tau} \left[ \dot{r}\left(L - \frac{2EMa}{r}\right)\mathbf{e}_\phi \right] \nonumber \\
    &= \left[ -\frac{L^2 \dot{r}}{r^2}\mathbf{e}_r + \left(\frac{L^2}{r} - M\right)\dot{\phi} \mathbf{e}_\phi \right] \nonumber \\
    &\quad - \left[ \left(\ddot{r}\left(L - \frac{2EMa}{r}\right) + \frac{2EMa\dot{r}^2}{r^2}\right)\mathbf{e}_\phi \right. \nonumber \\
    &\quad \left. - \dot{r}\left(L - \frac{2EMa}{r}\right)\dot{\phi} \mathbf{e}_r \right].
\end{align}
The radial components cancel to first order in $a$. We focus on the angular component $\dot{A}_\phi$:
\begin{equation}
    \dot{A}_\phi = \left(\frac{L^2}{r} - M\right)\dot \phi - \ddot{r}\left(L - \frac{2EMa}{r}\right) - \frac{2EMa\dot{r}^2}{r^2}.
\end{equation}
Substituting $\dot{\phi}$ from Eq. \eqref{eq:dot_phi_approx} and $\ddot{r}$ from Eq. \eqref{eq:ddotr}, and retaining terms only up to $\mathcal{O}(a)$:
Term 1:
\begin{align}
    &\left(\frac{L^2}{r} - M\right)\left(\frac{L}{r^2} + \frac{2EMa}{r^3}\right) \nonumber \\
    &\approx \frac{L^3}{r^3} - \frac{ML}{r^2} + \frac{2EML^2a}{r^4} - \frac{2M^2Ea}{r^3}.
\end{align}
Term 2
\begin{align}
    &-\left(-\frac{M}{r^2} + \frac{L^2}{r^3} - \frac{3ML^2}{r^4}\right)\left(L - \frac{2MEa}{r}\right) \nonumber \\
    &\approx \frac{ML}{r^2} - \frac{L^3}{r^3} + \frac{3ML^3}{r^4} - \frac{2M^2Ea}{r^3} + \frac{2L^2MEa}{r^4} - \frac{6M^2L^2Ea}{r^5}.
\end{align}
Term 3: Using the expression for $\dot{r}^2$ from Eq. \eqref{eq:dotr}:
\begin{align}
    \frac{2MEa}{r^2}\left(E^2 - 1 + \frac{2 M}{r} - \frac{L^2}{r^2} \right) = \nonumber \\ \frac{2MEa(E^2 - 1)}{r^2} + \frac{4 M^2Ea}{r^3} - \frac{2ML^2Ea}{r^4}
\end{align}
The summation of the above result yields
\begin{equation}
    \dot{A}_{\phi}= \frac{3ML^3}{r^4}-\frac{8 EM^2a}{r^3}-\frac{2MEa(E^2-1)}{r^2}+\frac{6M^2L^2Ea}{r^5}
\end{equation}

\bibliographystyle{apsrev4-2}
\bibliography{references}

\end{document}